\documentstyle[twoside,fleqn,espcrc2]{article}

\def\MeV{{\rm\  MeV}}
\def\GeV{{\rm\  GeV}}

\def\spose#1{\hbox to 0pt{#1\hss}}
\def\ltapprox{\mathrel{\spose{\lower 3pt\hbox{$\mathchar"218$}}
 \raise 2.0pt\hbox{$\mathchar"13C$}}}
\def\gtapprox{\mathrel{\spose{\lower 3pt\hbox{$\mathchar"218$}}
 \raise 2.0pt\hbox{$\mathchar"13E$}}}
\def\inapprox{\mathrel{\spose{\lower 3pt\hbox{$\mathchar"218$}}
 \raise 2.0pt\hbox{$\mathchar"232$}}}

\newcommand{\AmS}{{\protect\the\textfont2
  A\kern-.1667em\lower.5ex\hbox{M}\kern-.125emS}}

\title{Results from a study of the Nambu--Jona-Lasinio
model on the lattice}

\author{K.M. Bitar$\rm ^a$ and P. Vranas\thanks{speaker}
\address{Supercomputer Computations Research Institute,
Florida State University,\\
Tallahassee, Florida 32306--4052, USA} }

\begin{document}

\begin{abstract}

  The main results of our analysis of the two flavor
  Nambu--Jona-Lasinio model with $SU(2) \times SU(2)$ chiral symmetry
  on the four--dimensional hypercubic lattice with naive and Wilson
  fermions are presented. Large $N$ techniques and numerical
  simulations are used to study various properties of the model. The
  scalar and pseudoscalar spectrum, the approach to the continuum and
  chiral limits, the size of the $1/N$ corrections, and the effects of
  the zero momentum fermionic modes on finite lattices are studied.
  Also, some interesting observations are made by viewing the model as
  an embedding theory of the Higgs sector.

\end{abstract}

\maketitle

\section{Introduction}

When the high frequency modes of the gauge and fermionic fields of QCD
are integrated down to the energy scale $E$ corresponding the
correlation length of the gauge field ($E \approx$ glueball mass
$\approx 1550 \MeV$ \cite{UKQCD}) the resulting effective theory will
essentially be a theory of fermions with contact interactions and
cutoff $\Lambda \ltapprox 1550 \MeV$. The resulting effective
Lagrangian will maintain the original chiral symmetry but will be more
complicated. At energies much smaller than $\Lambda$ it should be
enough to keep in the Lagrangian the least irrelevant operator, namely
the four--Fermi dimension six operator. This is one way
\cite{Dhar-Shankar-Wadia} to motivate the Nambu--Jona-Lasinio (NJL)
model.

Unfortunately, by only keeping the four--Fermi operator, valuable
information is lost and the NJL model does not confine the quarks.
Furthermore, if, for example, we want to study the $\sigma$ particle,
which on phenomenological grounds is believed to have mass $\approx
750 \MeV$, then the separation of scales is probably not large enough
to justify the neglect of operators with dimension higher than six.
Nevertheless, the NJL model possesses the same chiral symmetry as QCD
and it can realize this symmetry in the Goldstone mode. It is this
feature that is most crucial in the understanding of the lightest
hadrons. Furthermore, our interest in the model is not so much aimed
at its quantitative predictability but rather on the qualitative
insights that can provide as a low energy theory of QCD, as an
embedding theory of the Scalar Sector of the Minimal Standard Model
and as a four--dimensional interacting theory of fermions on the
lattice.

The NJL model has been studied extensively for various cases with
continuum type regularizations. For a comprehensive review the reader
is referred to \cite{Klevansky} and references therein. Furthermore,
the NJL model is a special case of Yukawa models that, under a
different context, have been studied extensively with lattice
regularization \cite{Yukawa}. The model has also been studied on the
lattice \cite{Hasenfratz} in connection with the possible equivalence
of the top quark condensate with the Higgs field \cite{Bardeen}. Also
in this conference work was presented \cite{conf} regarding the study
of the NJL model on the lattice with staggered fermions in connection
with QED.

In this work \cite{Bitar-Vranas} we consider the two flavor (up and
down) NJL model with $SU(2) \times SU(2)$ chiral symmetry and SU(N)
color symmetry, with scalar and pseudoscalar couplings
\cite{Ebert-Volkov} on the four--dimensional hypercubic lattice and we
consider both naive and Wilson fermions. We study the NJL model using
large $N$ techniques and obtain analytical results both on finite and
infinite volumes. The infinite volume results are obtained
sufficiently close to the continuum limit using asymptotic expansions.
We also study the model for $N=2$ using an HMC numerical simulation
\cite{DKPR} with Conjugate Gradient and leap--frog algorithms.

\section{Results}

The seven main results that stem from our analysis are presented below.

\noindent {\bf 1) }
For naive fermions we calculate at large $N$ and with $M_\pi=140 \MeV$
the $\sigma$ mass ($M_\sigma$), the $\sigma$ width ($\Gamma_\sigma$)
and the constituent quark mass ($M_q$) in physical units as functions
of the cutoff. By setting $M_q = 310 \MeV$, we find $M_\sigma = 726
\MeV$, $\Gamma_\sigma=135 \MeV$, and $\Lambda = \pi / a = 1150 \MeV$.
$M_\sigma$ is consistent with phenomenological expectations and
$\Lambda$ is consistent with the expectation that the cutoff should be
close and below the mass of the lightest glueball ($1550 \MeV$). The
width however is underestimated. The reason is traced to the fact that
to leading order in large $N$ the width receives contributions only
from the quark bubble and not from the pion bubble because the pion
bubble is of order $1/N$. Because the phase space available for the
$\sigma$ to decay to two quarks is much smaller than the phase space
to decay to two pions the pion loop contribution, although of order
$1/N$, is probably more important than the quark loop contribution.

\noindent {\bf 2) }
The above result is relevant not only for the low energy QCD
but also for the Higgs sector. It is well known that there is an
equivalence between the $\sigma$-$\pi$ sector of QCD with the scalar
sector of the Minimal Standard Model. In the former the scale is set
by the pion decay constant ($F_\pi=93 \MeV$) and in the latter by the
weak scale ($F_\pi = 246 \GeV$). As mentioned above we find that in
accordance with phenomenological expectations in the $\sigma$-$\pi$
QCD sector $M_\sigma / F_\pi \approx 8$, but in the Higgs sector all
previous analysis predicts a triviality bound of the Higgs mass with
$M_\sigma / F_\pi \ltapprox 2.8$ (see for example \cite{HKNV}). In the
past this has been a reason for concern since it could imply that the
Higgs mass bound may be underestimated. Our analysis suggests that
this apparent discrepancy appears because the Higgs mass bounds are
traditionally obtained for $m_\sigma \ltapprox 0.5$ while the
$M_\sigma / F_\pi \approx 8$ ratio is obtained for $m_\sigma \approx
2$ and it should therefore be accompanied by large deviations from the
low energy behavior. Nevertheless, this is only a suggestion since we
have not calculated deviations from the low energy behavior of a
physical process that would enable us to make exact statements.
However, the value of the width serves as an indication of the size of
such deviations. In a way, departure from low energy behavior will be
signaled by an increasing width of the $\sigma$ to two quark decay.
At $m_\sigma \approx 2$ the width is already fairly large.

\noindent {\bf 3) }
If the Higgs sector is the low energy effective field theory
of a NJL model (which in turn is an effective theory of a high energy
QCD--like theory) with $N_c=3$, $n_f=2$ and $M_\pi=0$, then if we set
the fermion mass to ${M_q \over F_\pi} \approx {310 \over 93}$, as is
the case for the low energy sector of QCD, the Higgs mass will be
$M_\sigma = 1915\GeV$. This corresponds to $m_\sigma = 2$ and at this point
one would expect very large deviations from the low energy behavior of
scattering cross sections. This suggests that as the CM energy is
turned up, first the deviations from the low energy behavior will
become large signaling the onset of new physics, and later on the Higgs
particle would be observed.

\noindent {\bf 4) }
With Wilson fermions we obtain at large $N$ analytical expressions of
the pion mass ($m_\pi$) and constituent quark mass ($m_q$) in lattice
units as functions of the bare parameters of the model. We are then
able to make exact statements regarding the approach to the continuum
and chiral limits. We draw the ``phase diagram" and identify the
single point where a continuum chiral limit ($m_\pi \rightarrow 0$,
$m_q \rightarrow 0$) can be achieved.  This may provide an insight on
how the retrieval of the continuum chiral limit is achieved in QCD.

\noindent {\bf 5) }
At large $N$ and for Wilson fermions the $\sigma$ particle has mass
proportional to the cutoff. Our analysis traces this to two
related reasons. First, although the Wilson term has raised the masses
of the doublers to the cutoff, it has not decoupled them from the
theory. Through vacuum polarization these contribute to the $\sigma$
self energy and raise its mass. Second, although the Wilson term has
not altered the low frequency behavior of the propagating quark, it
has, however, altered its high frequency behavior. Again, through vacuum
polarization the high frequency modes contribute to the $\sigma$ self
energy and also raise its mass. Such a phenomenon may also be
responsible for the difficulty in observing a $\sigma$ particle in
numerical simulations of QCD with dynamical Wilson fermions.

\noindent {\bf 6) }
The numerical simulation is performed on finite lattices. For naive
fermions one would expect to be able to see some indication of the
chiral phase transition as well as a $\sigma$ particle. However, by
simply looking at the graph of the vacuum expectation value vs. the
coupling on an $8^3 \times 16$ lattice one can not see an
indication of a phase transition. Also the $\sigma$ particle in a
$16^4$ lattice is either non existent or too heavy to be measured.
Both of these unexpected results can be explained at large $N$. The
reason is traced to the existence of zero quark momentum modes that on
a finite lattice are not sufficiently suppressed. The zero modes
besides obscuring some of the physics are also probably partially
responsible for the large inversion times in the HMC algorithm. To
leading order at large $N$ the smallest eigenvalue of the matrix that
has to be inverted is $m_q^2$ and corresponds to the zero quark
momentum mode. For small $m_q$, the condition number of the matrix is
$4/m_q^2$ for $r=0$ and $64 r / m_q^2$ for $r=1$ and it is clear that
it depends strongly on the presence of the zero modes. A large
condition number will make the inversion of $M^\dagger M$ very slow.
Furthermore the spacing of the smallest eigenvalues behaves like $1 /
L^2$ and for larger lattices the inversion times will rapidly get
worse. An important observation can be made by noticing the dependence
of the condition number on $r$. This suggests that performing the
simulation with smaller $r$ will yield a quite faster inversion. It is
possible that this may also be the case for QCD.

\noindent {\bf 7) }
The observables measured in the numerical simulation (chiral
condensate, vacuum expectation value, pion wave function
renormalization constant, and pion mass) have values that are in good
agreement with leading order large $N$. This provides a quantitative
prediction for the size of the $1/N$ corrections. In agreement with
the large $N$ predictions discussed in {\bf (5)} and {\bf (6)} above,
the $\sigma$ mass was very heavy to give a good signal and was not
measured. Also, measurements of the sigma width were not performed,
but, as discussed in {\bf (1)} above, we expect the $1/N$ corrections
to the width to be large.

There are some interesting issues relevant to lattice work that have
not been considered in this paper.  It would be important to calculate
the three and four point vertices and therefore be able to calculate
scattering amplitudes and their departure from low energy behavior as
well as the $1/N$ corrections to the width. It would also be
interesting to study the NJL model at finite temperature. Finally, it
would be important to include vector meson couplings (see, for
example, \cite{Ebert-Volkov}, \cite{Ebert-Reinhard}) and confirm that
for the case of Wilson fermions the vector meson masses scale
appropriately and do not become of the order cutoff as the $\sigma$
particle does.

This research was supported in part by the DOE under grant $\#$
DE-FG05-85ER250000 and $\#$ DE-FG05-92ER40742.

\end{document}